\documentclass[pra,twocolumn]{revtex4}
\usepackage{graphicx}
\usepackage{amsmath}
\begin{document}
\title{Invariant Perturbation Theory of Adiabatic Process }

\author{Jian-Lan Chen, Mei-sheng Zhao, Jian-da Wu and Yong-de Zhang}
\address{Department of Mordern Physics and Hefei National Laboratory
                    for Physical Sciences at Microscale, University of Science and
                     Technology of China, Hefei, 230026,  China\ }

\begin{abstract}

In this paper we present an invariant perturbation theory of
adiabatic process according to the concepts of $U(1)$-invariant
adiabatic orbit and $U(1)$-invariant adiabatic expansion. The
probabilities of keeping the adiabatic orbit in the first-order and
the second-order approximation are calculated, respectively. We also
give a convenient sufficient condition.

\vspace{0.5cm}
 \noindent PACS number(s): 03.65.Ca, 03.65.Ta, 03.65.Vf
\end{abstract}

\maketitle


Quantum adiabatic theorem and adiabatic approximation are important
contents in quantum mechanics \cite{Ehrenfest,Born,Schwinger,Kato}.
Since the establishment of the quantum adiabatic theorem it has been
broadly applied both in theory and experiments. In the deep
investigation into quantum adiabatic process a lot of important
results have been obtained, for example, Landau-Zener Transition
\cite{Landau}, Gell-Mann-Low theorem \cite{Gell-Mann}, Berry
phase\cite{Berry}, holomony\cite{Simon}. Recently, the content of
quantum information and quantum computation has revived interest in
quantum adiabatic theorem and adiabatic approximation
\cite{Oreg,Schie,Zheng,Pillet,Farhi,Childs}. More recently, however,
the consistency of the traditional adiabatic approximation condition
has been doubted \cite{Marzlin,Tong1} and some new adiabatic
conditions are proposed \cite{Ye,Tong2,Yu}, which have caused
confusions. Then, it is important to find a proper and convenient
condition under which the evolution of the system can be considered
adiabatic.

This paper is aimed to propose a general invariant perturbation
theory for the quantum adiabatic process in terms of the concepts of
$U(1)$-invariant adiabatic orbit and $U(1)$-invariant adiabatic
expansion. Then, the probabilities of keeping in adiabatic orbit are
obtained in the first-order and the second-order approximation,
respectively. At last, a new and convenient adiabatic condition is
presented.

Firstly, we explain the invariant adiabatic basis and invariant
adiabatic expansion \cite{wu} with time-dependent coefficients. Let
us consider a time-dependent Hamiltonian $H(t)$ with the initial
state $|m,0 \rangle $ at time $t=0$, here $m$ denotes the initial
value of dimensionless quantum number set. We introduce a
dimensionless time parameter $\tau=E_{m}(0)t/\hbar$ and a
dimensionless Hamiltonian $h(\tau)=H(\tau)/E_{m}(0)$, $E_{m}(0) $ is
the energy of the initial state. The time-dependent
$Schr\ddot{o}dinger$ equation reads
\begin{eqnarray}
i\frac{\partial{|\Phi_{m}(\tau)\rangle}}{\partial{\tau}}&=&h(\tau)|\Phi_{m}(\tau)\rangle,\quad
|\Phi_{m}(\tau)\rangle|_{\tau=0}=|m,0\rangle  \nonumber\\
|\Phi_{m}(\tau)\rangle&=&Te^{-i\int_{0}^{\tau}h(\lambda)d\lambda}|m,0\rangle,
\label{g1}
\end{eqnarray}
here $T$ is time-ordered operator.

$Definition$ 1: A  state $\left| {\Phi _n \left( \tau \right)}
\right\rangle$, is a $dynamic$ $evolution$ $orbit$ of system when
the state $\left| {\Phi _n \left( \tau \right)} \right\rangle$
satisfies Eq.(\ref{g1}), describing an evolution orbit varying with
time in the Hilbert space.

Furthermore, if we consider $\tau$ as a fixed parameter, we can
always solve the following quasi-stationary state equation

\begin{equation}
h\left( \tau  \right)\left| {\varphi _n \left( \tau  \right)}
\right\rangle  = e_n \left( \tau  \right)\left| {\varphi _n \left(
\tau  \right)} \right\rangle \label{g2}
\end{equation}
then we will have the $adiabatic$ $solution$ or $adiabatic$ $orbit$
$\left| {\varphi _n \left( \tau  \right)} \right\rangle$ and its
corresponding eigenvalue $e_n \left( \tau  \right) = {{E_n \left(
\tau  \right)} \mathord{\left/ {\vphantom {{E_n \left( \tau \right)}
{E_n \left( 0 \right)}}} \right. \kern-\nulldelimiterspace} {E_m
\left( 0 \right)}}$.

Of course, although with the same initial state $|m,0\rangle$, the
dynamic evolution orbit $\left| {\Phi _m \left( \tau  \right)}
\right\rangle$ does not coincide with the adiabatic orbit $\left|
{\varphi _m \left( \tau \right)} \right\rangle$, or they are not
even close to each other. Furthermore, because of the Hermitian of
$h\left( \tau \right)$, all of these adiabatic orbits form a
complete basis of the system.

We denote $\gamma _{nm} \left( \tau  \right) \equiv i {\left\langle
{{\varphi _n (\lambda )}}
 \mathrel{\left | {\vphantom {{\varphi _n (\lambda )} {\dot \varphi _m (\lambda )}}}
 \right. \kern-\nulldelimiterspace}
 {{\dot \varphi _m (\lambda )}} \right\rangle }$ and the dot means the
 differentiation with respect to time. An adiabatic
 orbit multiplied by an arbitrary time-dependent phase factor still
 describes the same adiabatic orbit.

 $Definition$ 2: The following adiabatic orbits
\begin{equation}
\left| {\Phi _m^{adia} \left( \tau  \right)} \right\rangle  =
\exp\left\{ - i\int_0^\tau  {\left [e_m (\lambda ) - \gamma _{mm}
\left( \tau \right)\right]} d\lambda  \right\} \left| {\varphi _m
(\tau )} \right\rangle \label{g3}
\end{equation}
are invariant, up to a time-dependent phase factor, under $U(1)$
time-dependent transformation of the adiabatic orbit
\begin{eqnarray}
\left| {\varphi _m \left( \tau \right)}\right\rangle\to
e^{if_{m}\left( \tau \right)}\left| {\varphi _m \left( \tau \right)}
\right\rangle,   \label{g4}
\end{eqnarray}
with $f_m(0)=0$. We define this adiabatic orbit with special choice
of the time-dependent phase factor as $U(1)$ $invariant$ $adiabatic$
$basis$ which keeps the initial value of dimensionless quantum
number set $m$ invariant.

$Definition$ 3: If the $\left| {\varphi _m \left( \tau  \right)}
\right\rangle$ in Eq.(\ref{g3}) satisfies following conditions
\begin{equation}
\left\langle {{\varphi _n \left( \tau  \right)}}
 \mathrel{\left | {\vphantom {{\varphi _n \left( \tau  \right)} {\dot \varphi _m \left( \tau  \right)}}}
 \right. \kern-\nulldelimiterspace}
 {{\dot \varphi _m \left( \tau  \right)}} \right\rangle  = 0\,,\quad \forall n \ne m \label{g5}
\end{equation}
then ${\left| {\Phi _m^{adia} \left( \tau \right)} \right\rangle }$
is also the dynamic evolution solution of Eq.(\ref{g1}). We call
this adiabatic orbit "$adiabatic$ $evolution$ $orbit$" of system.

 Generally speaking, in an
arbitrary evolution process, the dynamic evolution orbit
$|\Phi_{m}(\tau)\rangle$  starting from the initial state
$|m,0\rangle$ will change or even vibrate rapidly among some
adiabatic orbits. This case can be described by the probability $P$
staying in the adiabatic obit
\begin{equation}
P_m\left( \tau  \right) = \left| {\left\langle {{\Phi _n^{adia}
\left( \tau  \right)}}
 \mathrel{\left | {\vphantom {{\Phi _n^{adia} \left( \tau  \right)} {\Phi _m \left( \tau  \right)}}}
 \right. \kern-\nulldelimiterspace}
 {{\Phi _m \left( \tau  \right)}} \right\rangle } \right|^2 ,\;\;\forall n \ne m \label{g6}
\end{equation}

Next our task is to find when the dynamic orbit is sufficiently
close to the adiabatic orbit if Eq.(\ref{g5}) is not satisfied.
Then, we will give the correct adiabatic approximation conditions.

In the $U(1)-invariant$ adiabatic basis the dynamic evolution orbit
reads
\begin{equation}
|\Phi_m (\tau)\rangle=\sum_n c_n (\tau)|\Phi_n
^{adia}(\tau)\rangle,\quad |\Phi_m (\tau)\rangle |_{\tau=0}
=|m,0\rangle, \label{g7}
\end{equation}
with initial conditions $c_m(0)=1,c_n(0)=0,\forall n\neq m$. The
time-dependent coefficients $c_k (\tau)$ are governed by
\begin{equation}
\dot{c}_n (\tau)=i\sum_{k\neq n}M(\tau)_{nk}c_k (\tau), \label{g8}
\end{equation}
where the diagonal elements of matrix $M(\tau)$ are zero and the
non-diagonal elements of $M(\tau)$ read
\begin{eqnarray}
M(\tau)_{k'k''}&=&\langle
\Phi_{k'}^{adia}(\tau)|i\frac{\partial}{\partial
\tau}|\Phi_{k''}^{adia}\rangle,\quad \forall k'\neq k''
\nonumber\\
&=& e^{i\alpha_{k'k''}(\lambda)}\left| {\gamma _{k'k''} \left(
\tau  \right)}, \right|\nonumber\\
\alpha_{k'k''} (\tau)&=&\int_{0}^{\tau}d\eta \left (e_{k'}
(\eta)-e_{k''} (\eta)\right )+\xi_{k'k''}(\eta),
 \label{g9}
\end{eqnarray}
here $ \xi _{mn} \left( \tau \right) \equiv \int_0^\tau {d\eta
\,\left( {\gamma _{nn} \left( \eta  \right) - \gamma _{mm} \left(
\eta \right)} \right)} + \arg \gamma _{mn} \left( \tau \right)$.
${\Delta} _{mn}=\dot {\xi} _{mn}$ is referred as $ geometric $
$potential$ of this system. And for $ geometric $ $potential$ one
can obtain further detailed analysis and application in our other
papers \cite{zhao}.

Then we can get the expanding coefficients with initial conditions
$\vec{C}(0) \left(c_m (0)=1,c_k (0)=0,\forall k\neq m\right)$
\begin{equation}
\vec{C}(\tau)=\left ( T\exp\left [
i\int_{0}^{\tau}M(\lambda)d\lambda\right ]\right
)\vec{C}(0).\label{g10}
\end{equation}
The element of Eq.(\ref{g10}) is
\begin{equation}
c_k (\tau)=\left ( T\exp\left [
i\int_{0}^{\tau}M(\lambda)d\lambda\right ]\right )_{km}.\label{g11}
\end{equation}
Apparently, Eq.(\ref{g11}) shows that the dynamic evolution is just
an adiabatic evolution if Eq.(\ref{g5}) is satisfied. In addition,
since $M(\tau)$ is Hermitian, the probability of the evolution is
conservative, that is
\begin{equation}
|c_m(\tau)|^2+\sum_{k}|c_k (\tau)|^2=1, \label{g12}
\end{equation}
which shows the time-dependent system considered is not a
dissipative system.

Secondly, we try to get the probability $P_m (\tau)$ of keeping in
adiabatic orbit $|\phi_m (\tau)\rangle$. In the time-dependent
dynamic evolution process, the probability of keeping in the
adiabatic orbit $|\Phi_m ^{adia}(\tau)\rangle$, i.e., keeping the
dimensionless quantum numbers invariant, is
\begin{equation}
P_m (\tau)=|c_m (\tau)|^2=\left |\left (T\exp\left
[i\int_{0}^{\tau}M(\lambda)\right ]\right)_{mm}\right
|^2,\label{g13}
\end{equation}
Then adiabatic approximation requires
\begin{equation}
P_m (\tau)\to 1  ,\label{g14}
\end{equation}
It means that the transition probability from dynamic evolution
orbit to any other adiabatic orbits can be neglected.

We can get the analytical expression of $P_m (\tau) $ using the
approach of coefficients ratio. Integrate Eq.(\ref{g8}) and get
\begin{equation}
c_n(\tau)=\prod_{k\neq n}\exp\left
[i\int_{o}^{\tau}M(\lambda)_{nk}\frac{c_k (\tau)}{c_m (\tau)}\right
]\quad \forall n. \label{g15}
\end{equation}
Obviously, this equation set can be solved by the iterative method.
Consider Eq.(\ref{g11}) we can get the probability of remaining in
$|\Phi_m ^{adia}(\tau)\rangle$ from Eq.(\ref{g15})
\begin{widetext}
\begin{eqnarray}
 P_m \left( \tau  \right)
 =\prod\limits_{k \ne m} {\left| {\exp \left\{ {i\int_0^\tau {d\lambda
} \frac{{\left\{ {T\exp \left[ {i\int_0^\lambda  {d\eta } M\left(
\eta \right)} \right]} \right\}_{km} }}{{\left\{ {T\exp \left[
{i\int_0^\lambda {d\eta M\left( \eta  \right)} } \right]}
\right\}_{mm} }}M\left( \lambda \right)_{mk} } \right\}} \right|}
^2.\nonumber\\
\label{g16}
\end{eqnarray}
\end{widetext}
From Eq.(\ref{g16}) we can also obtain a necessary and sufficient
condition in a compact form
\begin{widetext}
\begin{equation}
{\mathop{\rm Re}\nolimits} \left\{ {i\sum\limits_{k \ne m}
{\int_0^\tau  {d\lambda } \frac{{\left( {T\exp \left[
{i\int_0^\lambda  {d\eta } \,M\left( \eta  \right)} \right]}
\right)_{km} }}{{\left( {T\exp \left[ {i\int_0^\lambda  {d\eta }
\,M\left( \eta  \right)} \right]} \right)_{mm} }}M_{mk} \left(
\lambda  \right)} } \right\} \to 0. \label{g17}
\end{equation}
\end{widetext}

Thirdly, we give various approximate approaches to calculate the
probability.

\noindent 1, According to Eq.(\ref{g12}) the first-order
approximation of $P_m (\tau)$ is
\begin{equation}
P_m (\tau)\cong  1-\sum_{k\neq m }\left |
\int_{0}^{\tau}d\lambda\langle \Phi_k ^{adi}(\lambda)| {\dot{\Phi}_m
^{adi}(\lambda)\rangle} \right| .\label{g18}
\end{equation}
Then the necessary and sufficient conditions can be described as
\begin{equation}
\left | \int_{0}^{\tau}d\lambda \langle \Phi_k
^{adi}(\lambda)|\dot{\Phi}_m ^{adi}(\lambda)\rangle \right |^2\to
0,\quad \forall k\neq m. \label{g19}
\end{equation}
Eq.(\ref{g19}) is of abundant contents which has been discussed
 in another paper \cite{wu}. Later we will obtain a
convenient adiabatic condition from Eq.(\ref{g19}).

\noindent 2, From Eq.(\ref{g11}) the second-order approximation of
$P_m (\tau)$ is
\begin{equation}
P_m (\tau)\cong
  \left| {\left( {1 - \int_0^\tau  {d\lambda _1 \int_0^{\lambda _1 }
 {d\lambda _2 M\left( {\lambda _1 } \right)M\left( {\lambda _2 } \right)} } }
  \right)_{mm} } \right|^2.
 \label{g20}
\end{equation}
because the first-order term of Eq.(\ref{g11}) is zero. The
adiabatic approximate reads
\begin{equation}
\left| {\sum\limits_{k \ne m} {\int_0^\tau  {d\lambda _1
\int_0^{\lambda _1 } {d\lambda _2 M_{mk} \left( {\lambda _1 }
\right)M_{km} \left( {\lambda _2 } \right)} } } } \right|^2  \to 0.
 \label{g21}
\end{equation}

\noindent 3, the first-order of the coefficients ratio is. $c_k
(\tau)$, $c_m (\tau)$ are all approximated in the first-order with
$c_m (\tau)\cong 1$
\begin{equation}
\frac{c_k (\tau)}{c_m (\tau)}\cong i\int_{0}^{\tau}d\eta
M_{km}(\eta).\label{g22}
\end{equation}
We substitute Eq.(\ref{g22}) into Eq.(\ref{g17})
\begin{equation}
{\mathop{\rm Re}\nolimits} \left\{ { - \sum\limits_{k \ne m}
{\int_0^\tau  {d\lambda } M_{mk} \left( \lambda
\right)\int_0^\lambda  {d\eta } \,M_{km} \left( \eta  \right)} }
\right\} \to 0 .\label{g23}
\end{equation}
It should be pointed out that this method can only be applied to the
situation without reversion of quantum state which can not be
treated as adiabatic process with $ c_m (\tau')=0$

Fourthly, as application of adiabatic invariant perturbation theory
we consider a spin-1/2 charged particle in magnetic field.
Eq.(\ref{g19}) can be rewritten as
\begin{equation}
\left |\int_0 ^{\tau}d\lambda M_{km}\right |^2= \left |\int_0
^{\tau}d\lambda e^{i\alpha(\lambda)}|\gamma_{km}|\right |^2\to
0,\quad \forall k\neq m \label{g24}
\end{equation}
 Now we suppose
$\ddot{\alpha}(\tau)=0$, then $\alpha(\tau)$ can be linearly
expanded
\begin{equation}
e^{i\alpha(\tau )}  = e^{i\left( {\alpha_0  + \Omega _0 \tau }
\right)} .\label{g25}
\end{equation}
where $\alpha_0$,  $\Omega_0 $ are constants. $\left| {\gamma _{km}
\left( \lambda  \right)} \right|$ is a periodic function can be
expanded in Fourier series
\begin{eqnarray}
\left| {\int_0^\tau  {d\lambda e^{i\alpha (\lambda )} \left| {\gamma
_{km} \left( \lambda  \right)} \right|} } \right|^2  &=& \left|
{\int_0^\tau  {d\lambda e^{i\Omega _0 \lambda } \sum\limits_l
{\Gamma _l^{\left( {km} \right)} e^{i\Omega _{km.l} \,\lambda } } }
} \right|^2 \nonumber\\
&=&\left| {\sum\limits_l {\frac{{\Gamma _l^{\left( {km} \right)}
}}{{\left( {\Omega _0  + \Omega _{km,\,l} } \right)}}} } \right|^2.
\label{g26}
\end{eqnarray}
 The adiabatic approximate condition reads
 \begin{equation}
\left| {{\frac{{\Gamma _l^{\left( {km} \right)} }}{{\left( {\Omega
_0  + \Omega _{km,\,l} } \right)}}} }\right|  \ll 1,\quad
\ddot{\alpha}(\tau)=0,  \label{g27}
\end{equation}
 which is a convenient sufficient conditions \cite{cjl}.
In fact, $\ddot{\alpha}(\tau)=0$ is a physical requirement. For this
kind of Hamiltonian \cite{Yu}
 \begin{equation}
 H_V(t)=e^{-itV}H e^{itV}   ,\label{g28}
 \end{equation}
 where $V$ and $H=\sum_{n} E_n|E_n\rangle\langle E_n| $ are two
 arbitrary time-independent Hamiltonian with $\{E_n,|E_n\rangle\}$
 being the eigensystem of $H$. The quasi-stationary state of
 $H_V(t)$ is $e^{-itV}|E_n\rangle$.
 The adiabatic orbit reads
 \begin{equation}
 |\Phi_m^{adi}(\tau)\rangle  =e^{-iE_m\tau +i\langle E_m |V|E_m\rangle
 \tau}e^{-itV}|E_n\rangle.\label{g30}
 \end{equation}
Then we get the elements of the matrix $M$
\begin{eqnarray}
{M}_{nm}&=&\langle
\Phi_n^{adi}(\tau)|\dot{\Phi}_m^{adi}(\tau)\rangle
\nonumber\\
&=&e^{-i(E_m-E_n) \tau  +i\langle E_m|V|E_m\rangle \tau  -i\langle
E_n|V|E_n\rangle \tau}  \langle E_n|V|E_m\rangle
.\nonumber\\
\label{g31}
\end{eqnarray}
The terms about $\tau$ in the exponent are all linear, which means
$\ddot{\alpha}=0$ is satisfied. Therefore, we can prove concisely
 that the adiabatic condition Eq.(\ref{g27}) is sufficient for this kind of Hamiltonian.
  Next we will explain our above adiabatic condition Eq.(\ref{g27})
with model in \cite{Tong1}. For system $a$, the adiabatic
approximate conditions is $\left | \frac{\omega \sin\theta}{\omega_0
+\omega\cos\theta}\right |\ll 1$, under this condition the
probability of remaining in adiabatic orbit is
$P_m=1-\frac{\omega^2\sin^2\theta}{\tilde{\omega}^2}\sin^2\frac{\tilde{\omega}t}{2}\to
1$; For system $b$, the adiabatic approximate conditions is $\left |
\tan\theta\right |\ll 1$, under this condition the probability of
remaining in adiabatic orbit is
$P_m=1-\sin^2\theta\sin^2\frac{{\omega}t}{2}\to 1$. It is not
difficult to see that the adiabatic conditions for system $a$ and
system $b$ are different, then the inconsistency showed by this
example does not exist when applying our condition. The trivial and
special example of our condition with constant geometric potential
and energy gap is proved in \cite{Tong2}. Of course, our condition
with time-dependent geometric potential is more natural and general.

In conclusion, we present the invariant perturbation theory based on
the concepts of invariant adiabatic orbit and expansion stated in
our paper. The probability of keeping int the adiabatic orbit is
given. Furthermore, we give an convenient adiabatic approximation
condition which is more convenient to apply than the general
sufficient showed in \cite{Tong1}. The derivation of our condition
is more concise and general than \cite{Yu}. The result is that we
can obtain the approximate dynamic solution in perturbation theory
by using the corresponding quasi-stationary equations and energy for
nondissipative and smooth time-dependent process.

\acknowledgments

We thank Prof. Si-xia Yu for illuminating discussion. This work is
supported by the NNSF of China, the CAS, and the National
Fundamental Research Program (under Grant No. 2006CB921900).


\begin{thebibliography}{99}
\bibitem{cjl}         It is easy to prove that the condition Eq.(\ref{g27}) is
                      sufficient because this condition can guarantee that every order of iterative process
                      is convergent.
\end{thebibliography}
\end{document}